\documentclass[12pt]{article}
\usepackage{color}
\usepackage{amssymb}
\usepackage[centertags]{amsmath}
\usepackage[small,loose]{subfigure}  % subfigures with (a), (b) etc., ... and own caption
\usepackage{graphicx}
\usepackage{bbm}
\usepackage[small]{caption2} % font of captions is different from text

\DeclareMathOperator{\diag}{diag}

\setcaptionwidth{0.75\textwidth}

\renewcommand{\thefootnote}{\fnsymbol{footnote}}
\def\I{\ensuremath{\mathrm{i}}}

\begin{document}

\title{
\begin{flushright}
\ \\*[-80pt]
\begin{minipage}{0.3\linewidth}
\normalsize
%hep-th/0611020 \\
KUNS-2044 \\
OHSTPY-HEP-T-06-005 \\
NSF-KITP-06-99 \\
TUM-HEP-650/06 \\*[50pt]
\end{minipage}
\end{flushright}
{\Huge \bf Stringy origin of non-Abelian discrete flavor symmetries
\\*[20pt]}}

\author{Tatsuo~Kobayashi$^{1}$, \
Hans~Peter~Nilles$^{2}$, \
Felix~Pl\"oger$^{2}$, \ \\
Stuart Raby$^{3}$ \
and \
Michael~Ratz$^{4}$ \\*[20pt]
$^1${\it \normalsize
Department of Physics, Kyoto University,
Kyoto 606-8502, Japan} \\[0.2cm]
$^2${\it \normalsize Physikalisches Institut,
Universit\"at Bonn, Nussalle 12, D-53115 Bonn, Germany}\\[0.2cm]
$^3${\it \normalsize Department of Physics, The Ohio State University, Columbus,
OH 43210 USA}\\[0.2cm]
$^4${\it \normalsize Physik Department T30,
Technische Universit\"at M\"unchen, D-85748 Garching, Germany}
\\*[50pt]}

\date{
\centerline{\small \bf Abstract}
\begin{minipage}{0.9\linewidth}
\medskip
\medskip
%\small
We study the origin of non-Abelian discrete flavor symmetries in
superstring theory. We classify all possible non-Abelian discrete
flavor symmetries which can appear in heterotic orbifold models.
These symmetries include $D_4$ and $\Delta(54)$. We find that the
symmetries of the couplings are always larger than the symmetries of
the compact space.  This is because they are a
consequence of the geometry of the orbifold combined  with the space
group selection rules of the string. We also study possible
breaking patterns. Our analysis yields a simple geometric
understanding of the realization of non-Abelian flavor symmetries.
\end{minipage}
}

\begin{titlepage}
\maketitle
\thispagestyle{empty}
\end{titlepage}

%\tableofcontents

\renewcommand{\thefootnote}{\arabic{footnote}}
\setcounter{footnote}{0}

\section{Introduction}

One of the most important issues in contemporary particle physics is to
understand the quark and lepton flavor structure,  i.e.\ the origin of the
number of generations, the observed mass hierarchies as well as the mixing
angles. Many attempts to understand flavor are based on spontaneously broken
Abelian \cite{Froggatt:1978nt} and non-Abelian flavor symmetries
\cite{Frampton:1994rk}, such as
$S_3(\approx D_3)$~\cite{flavor-s3},
$S_4$~\cite{flavor-S4},
$A_4$~\cite{flavor-A4},
$D_4$~\cite{flavor-D4},
$D_5$~\cite{Hagedorn:2006ir},
$Q_6$~\cite{Babu:2004tn},
$\Delta$-subgroups of SU(3)~\cite{Kaplan:1993ej,Chou:1995di,deMedeirosVarzielas:2006fc}, governing Yukawa couplings for quarks
and leptons. Discrete symmetries are not only useful to understand flavor issues
(e.g.\ the observed large mixing angles in the lepton sector) but also to
control soft supersymmetry breaking terms, in particular to suppress dangerous
flavor changing neutral currents. However, the origin of these discrete
symmetries remains obscure in the framework of 4D field theory.

It is not surprising that compactifications of higher-dimensional field theories
offer an explanation for the appearance of non-Abelian discrete flavor
symmetries,  because the latter are symmetries of certain geometrical solids. The
symmetries of internal space give rise to symmetries of the interactions between
localized fields, which may eventually become flavor symmetries. To be able to
evaluate the couplings, and to identify their symmetries, requires the
specification of a framework.

We base our analysis on superstring theory, which is a
promising candidate for a unified description of nature, including
gravity. Consistent (super-)string theories have, in addition to 4D
Minkowski space-time, six extra dimensions. An important aspect of
string compactifications is that phenomenological features, such as
the number of generations and the structure of Yukawa couplings, are
determined by geometrical properties of the 6D compact space. 4D
string models often enjoy Abelian discrete $\mathbbm{Z}_N$ symmetries,
which govern the allowed couplings. On the other
hand, non-Abelian discrete flavor symmetries derived from string
theory have not yet been studied extensively in the literature.
The purpose of the present study is to fill this gap.

Among the known string constructions, heterotic orbifold models
\cite{Dixon,IMNQ} have a particularly simple geometric
interpretation, and an encouraging phenomenology. The  selection
rules for heterotic orbifolds are well known
\cite{Hamidi:1986vh,Dixon:1986qv,Kobayashi:1991rp}, but little
attention has been paid to the emerging non-Abelian flavor
symmetries.  Recently, explicit string compactifications, based on
the $\mathbbm{Z}_6\mathrm{-II}=\mathbbm{Z}_2 \times \mathbbm{Z}_3$
heterotic orbifold, with a $D_4$ flavor symmetry have been
constructed
\cite{Kobayashi:2004ud,Kobayashi:2004ya,Buchmuller:2005jr,Buchmuller:2006ik}.
In these models, the three generations are comprised of a singlet
and a doublet under the $D_4$ symmetry. This $D_4$ flavor symmetry
has important phenomenological implications
\cite{Kobayashi:2004ya,next}.

In this paper we study  which types of non-Abelian discrete flavor symmetries
can appear in heterotic orbifold models. We classify all the possible
non-Abelian discrete symmetries which can arise from heterotic orbifold models,
and explore which representations appear in the zero-modes.

The paper is organized as follows.
In section 2, we collect some basic facts on strings on orbifolds.
Sections 3 and 4 are dedicated to a classification of all
the possible non-Abelian discrete flavor symmetries.
Their breaking patterns are discussed in section 5.
Section 6 is devoted to conclusions and discussion.
In appendix \ref{app:Couplings} we outline the calculation of coupling strengths
on orbifolds.
The appendices \ref{app:D4} and \ref{app:Delta54} deal with group-theoretical
aspects of $D_4$ and $\Delta(54)$.

\section{Strings on orbifolds}

\subsection{Review of basic facts}

Let us start with a brief introduction to strings on orbifolds
\cite{Dixon,Katsuki:1989bf} (for recent reviews see
\cite{Forste:2004ie,Kobayashi:2004ya,Buchmuller:2006ik}). A $d$-dimensional
orbifold emerges by dividing a $d$-dimensional torus $\mathbbm{T}^d$ by its
symmetry, represented by an automorphism (`twist') $\theta$.  $\mathbbm{T}^d$ is
obtained as $\mathbbm{R}^d/\Lambda$, where  $\Lambda$ is a $d$-dimensional
lattice, and the twist $\theta$ is the finite-order automorphism of $\Lambda$,
i.e.\ $\theta\,\Lambda=\Lambda$ and $\theta^N=\mathbbm{1}$. The orbifold is then
denoted as $\mathbbm{T}^d/\mathbbm{Z}_N$.  In other words, $\mathbbm{T}^d$
emerges from $\mathbbm{R}^d$ through the identification
\begin{equation}
 x^i~ \sim~ x^i + n_a\, e^i_a\;,
\end{equation}
where $n_a$ is integer and $\{e^i_a\}$ is the lattice basis of $\Lambda$.
Furthermore, one identifies points related by $\theta$,
\begin{equation}
 (\theta x)^i~\sim~x^i + n_a\, e^i_a\;,
\end{equation}
on the orbifold. We are specifically interested in 6D orbifolds which preserve
N=1 supersymmetry in 4D. For those it is convenient to diagonalize the twist
$\theta$, i.e.\ parametrize $\mathbbm{T}^6$ by three complex coordinates $z^i$
w.r.t.\ which
\begin{equation}\label{eq:TwistDiag}
 \theta
 ~=~
 \diag \bigl(\exp(2\pi\I\,v_1),\exp(2\pi\I\,v_2),\exp(2\pi\I\,v_3)\bigr)\;.
\end{equation}
Hereby $v_i=n_i/N$ ($n_i\in\mathbbm{Z}$) and $\sum_iv_i=0$.

Among the Abelian orbifolds, only certain constructions lead to N=1
supersymmetry in 4D. There are, first of all, nine classes of $\mathbbm{Z}_N$
orbifolds \cite{Dixon} which are surveyed in table \ref{tab-0} (a). Here the
second column shows  $v_i$ as introduced in \eqref{eq:TwistDiag}. Moreover,
there are $\mathbbm{Z}_N \times \mathbbm{Z}_M$ orbifolds which have two
independent twists, $\theta$ and $\omega$ with  $\theta^N=\mathbbm{1}$ and
$\omega^M=\mathbbm{1}$ (cf.\ \cite{Font:1988mk}). Nine classes of $\mathbbm{Z}_N
\times \mathbbm{Z}_M$ orbifold models  lead to N=1 SUSY. Their twists are shown
in table \ref{tab-0} (b) as  $\theta = \diag (e^{2 \pi\I v^1_1}, e^{2
\pi\I v^1_2}, e^{2 \pi\I v^1_3})$ and  $\omega = \diag  (e^{2 \pi\I v^2_1},
e^{2 \pi\I v^2_2}, e^{2 \pi\I v^2_3})$. Note that the $\mathbbm{Z}_2 \times
\mathbbm{Z}_3$ orbifold is equivalent  to the $\mathbbm{Z}_6$-II orbifold.

\begin{table}[!h]
\centerline{%
\subtable[$\mathbbm{Z}_N$. \label{tab-1}]{
\begin{tabular}{|c|c|} \hline
orbifold & twist \\  \hline \hline
$\mathbbm{Z}_3$ & $(1,1,-2)/3$  \\
$\mathbbm{Z}_4$ & $(1,1,-2)/4$  \\
$\mathbbm{Z}_6$-I & $(1,1,-2)/6$  \\
$\mathbbm{Z}_6$-II & $(1,2,-3)/6$  \\
$\mathbbm{Z}_7$ & $(1,2,-3)/7$ \\
$\mathbbm{Z}_8$-I & $(1,2,-3)/8$  \\
$\mathbbm{Z}_8$-II & $(1,3,-4)/8$  \\
$\mathbbm{Z}_{12}$-I & $(1,4,-5)/12$  \\
$\mathbbm{Z}_{12}$-II & $(1,5,-6)/12$  \\ \hline
\end{tabular}}
\quad
\subtable[$\mathbbm{Z}_N\times\mathbbm{Z}_M$.\label{tab-2}]{
\begin{tabular}{|c|c|c|} \hline
orbifold & $v^1$ & $v^2$\\  \hline \hline
$\mathbbm{Z}_2 \times \mathbbm{Z}_2$ & $(1,0,-1)/2$  &  $(0,1,-1)/2$\\
$\mathbbm{Z}_2 \times \mathbbm{Z}_3$ & $(1,0,-1)/2$  &  $(0,1,-1)/3$\\
$\mathbbm{Z}_2 \times \mathbbm{Z}_4$ & $(1,0,-1)/2$  &  $(0,1,-1)/4$\\
$\mathbbm{Z}_2 \times \mathbbm{Z}_6$ & $(1,0,-1)/2$  &  $(0,1,-1)/6$\\
$\mathbbm{Z}_2 \times \mathbbm{Z}'_6$ & $(1,0,-1)/2$  &  $(1,1,-2)/6$\\
$\mathbbm{Z}_3 \times \mathbbm{Z}_3$ & $(1,0,-1)/3$  &  $(0,1,-1)/3$\\
$\mathbbm{Z}_3 \times \mathbbm{Z}_6$ & $(1,0,-1)/3$  &  $(0,1,-1)/6$\\
$\mathbbm{Z}_4 \times \mathbbm{Z}_4$ & $(1,0,-1)/4$  &  $(0,1,-1)/4$\\
$\mathbbm{Z}_6 \times \mathbbm{Z}_6$ & $(1,0,-1)/6$  &  $(0,1,-1)/6$\\
\hline
\end{tabular}
}
}
\caption{(a) $\mathbbm{Z}_N$ and (b) $\mathbbm{Z}_N\times\mathbbm{Z}_M$ orbifold
twists for 6D $\mathbbm{Z}_N$ orbifolds leading to N=1 SUSY.}
\label{tab-0}
\end{table}

Zero-modes, described by string coordinates $X^i$, of an orbifold arise from
closed strings, satisfying the  boundary conditions
\begin{equation}\label{eq:BoundaryCondition}
 X^i(\sigma + \pi )~=~(\theta^k X)^i(\sigma) + n_a\, e^i_a\;.
\end{equation}
These boundary conditions can be either untwisted, i.e.\  $k=0$, or twisted,
i.e.\ $1\le k \le N-1$. Correspondingly, the Hilbert space of (massless) states
decomposes in an untwisted and various twisted sectors, denoted by $U$
and $T_k$, respectively.
The states from the untwisted sector are bulk fields in the effective field
theory whereas the twisted states are brane fields living at the fixed points or
planes.
More specifically, the center-of-mass coordinates of twisted sector zero-modes
satisfy an analogous condition to \eqref{eq:BoundaryCondition},
\begin{equation}
x^i~=~(\theta^k x)^i + n_a\, e^i_a\;,
\end{equation}
and are therefore in one-to-one correspondence to the  fixed points or fixed
planes of the orbifold. It is common to denote the fixed points or planes by the 
corresponding space group element $(\theta^k,n_a e^i_a)$. The product of two
space group elements,  $(\theta^{k^{(1)}}, \ell^{(1)})$ and $(\theta^{k^{(2)}},
\ell^{(2)})$, is defined by
\begin{equation}
(\theta^{k^{(1)}}, \ell^{(1)})\, (\theta^{k^{(2)}},\ell^{(2)})~=~
(\theta^{k^{(1)}+k^{(2)}}, \theta^{k^{(1)}} \ell^{(2)} +\ell^{(1)})\;.
\end{equation}
Since the orbifold identification implies $\Lambda \sim \theta^k \Lambda$, space
group elements $(\theta^k,\ell)$ are only defined up to translations in the
sublattice $\Lambda_k=(\mathbbm{1} - \theta^k)\,\Lambda$, i.e.\
$(\theta,\ell)\simeq(\theta,\ell+(\mathbbm{1} - \theta^k)\,\lambda)$ with
$\lambda\in\Lambda$. In other words, the space group elements appear in
conjugacy classes $(\theta^k,\ell + (\mathbbm{1} - \theta^k)\,\Lambda)$, and each
conjugacy class corresponds to an independent fixed point or plane.

\subsection{Couplings on orbifolds}

Unlike in the field-theoretic case, coupling strengths are not free
parameters in string theory but calculable. In what follows, we give a brief
review on Yukawa couplings as well as $n$-point couplings
\cite{Hamidi:1986vh,Dixon:1986qv,Burwick:1990tu}.

First of all, couplings in heterotic orbifolds are dictated by selection rules
\cite{Hamidi:1986vh,Dixon:1986qv,Kobayashi:1991rp,Kobayashi:2004ya,Buchmuller:2006ik}. Apart
from gauge invariance and $H$-momentum conservation, allowed couplings are
subject to the space group selection rules. An $n$-point coupling among string
states  corresponding to fixed points $(\theta^{k^{(j)}}, \ell^{(j)})$
$(j=1,\cdots,n)$ is allowed only if their product includes the identity,
\begin{equation}\label{eq:SpaceGroupRule}
 \prod_{j=1}^n (\theta^{k^{(j)}}, \ell^{(j)})~\simeq~
 \bigl(\mathbbm{1}, 0\bigr)
 \;.
\end{equation}

Moreover, the coupling strength between localized fields is a function of
geometrical features such as the distance between the fields. We review the
computation of coupling strengths in appendix \ref{app:Couplings}. The important
fact for the subsequent discussion is that, if the geometrical settings of two
couplings coincide, the coupling strengths coincide as well. In the next
sections we shall study the implications of this statement and the space group
rule \eqref{eq:SpaceGroupRule}.

\section{Non-Abelian flavor symmetries of building blocks}
\label{sec:nonAbelianSymmetries}

In many cases, the torus $\mathbbm{T}^6$ factorizes in tori of smaller
dimensions, i.e.\ the lattice $\Lambda$ decomposes in orthogonal sublattices.
One is then lead to consider the building blocks
\begin{equation}
 \mathbbm{S}^1/\mathbbm{Z}_2\;, \quad \mathbbm{T}^2/\mathbbm{Z}_3\;, \
 \quad \mathbbm{T}^2/\mathbbm{Z}_4\;, \quad \mathbbm{T}^2/\mathbbm{Z}_6\;,
 \quad \mathbbm{T}^4/\mathbbm{Z}_8\;,
 \quad \mathbbm{T}^4/\mathbbm{Z}_{12}\;,
 \quad \mathbbm{T}^6/\mathbbm{Z}_7\;,
%\label{factor-orbi}
 \label{eq:BuildingBlocks}
\end{equation}
which arise from the 6D orbifold by projection.\footnote{%
One should, however, use these building blocks with caution. For instance, the
$\mathbbm{Z}_6$-II orbifold based on the root lattice of
$\mathrm{G}_2\times\mathrm{SU}(3)\times\mathrm{SO}(4)$
\cite{Kobayashi:2004ud,Kobayashi:2004ya,Buchmuller:2004hv,Buchmuller:2006ik},
$(\mathbbm{T}^2_{\mathrm{G}_2}\times\mathbbm{T}^2_{\mathrm{SU}(3)}\times
\mathbbm{T}^2_{\mathrm{SO}(4)})/\mathbbm{Z}_6$, is clearly not equivalent to
$(\mathbbm{T}^2_{\mathrm{G}_2}/\mathbbm{Z}_6)\times(\mathbbm{T}^2_{\mathrm{SU}(3)}/\mathbbm{Z}_3)\times
(\mathbbm{T}^2_{\mathrm{SO}(4)}/\mathbbm{Z}_2)$ since the twist $\theta$
% (as well as the subtwists $\theta^2,\,\theta^3$)
acts on the three two-tori simultaneously.}
These building blocks play an important role when discussing orbifold GUT limits
\cite{Kobayashi:2004ud,Forste:2004ie,Kobayashi:2004ya,Buchmuller:2004hv,Buchmuller:2006ik},
where one considers the effective field theory describing anisotropic orbifolds
for energies between different compactification scales.

An important property of the space group rule
\eqref{eq:SpaceGroupRule} is that, if the torus factorizes, it can
be expressed in terms of independent sub-conditions that have to be
fulfilled separately for the building blocks. In what follows, we
will explain this statement in more detail and study the
consequences of rule \eqref{eq:SpaceGroupRule}. The discrete flavor
symmetries of combinations of building blocks will be studied in
section \ref{sec:CombiningBuildingBlocks}.

\subsection{$\boldsymbol{\mathbbm{S}^1/\mathbbm{Z}_2}$ orbifold}
\label{3-1}

\begin{figure}[h]
\centerline{\includegraphics{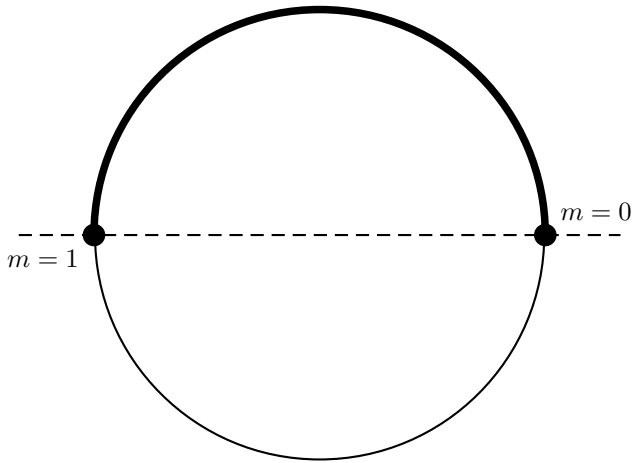}}
\caption{$\mathbbm{S}^1/\mathbbm{Z}_2$ orbifold. Points which are related by a
reflection on the dashed line are identified. The fundamental region of the
orbifold is an interval with the fixed points sitting at the boundaries.}
\label{fig:S1overZ2}
\end{figure}

In the 1D orbicircle, i.e.\ the $\mathbbm{S}^1/\mathbbm{Z}_2$ orbifold
(figure~\ref{fig:S1overZ2}),  there are two independent fixed points, which are
denoted by  their corresponding space group elements,
\begin{equation}
(\theta, m\, e)\;.
\end{equation}
Here $\theta$ is the $\mathbbm{Z}_2$ twist (reflection), $m=0,1$ and the $e$ is
the unit vector defining $\mathbbm{S}^1$, i.e.\ we identify $x \sim x +e$ on
$\mathbbm{R}^1$. The sublattice $(\mathbbm{1} - \theta)\Lambda$ is spanned by
$2e$. That implies that there are two conjugacy classes corresponding to $
(\theta, m\, e)$ with $m$ odd and even. The above space group elements with
$m=0,1$  are their representatives.

Consequently, there are two types of twisted strings corresponding to the above
independent fixed points $(\theta, m\, e)$ with  $m=0,1$
(figure~\ref{fig:S1overZ2}). In the field-theoretic description, these are brane
matter fields living on these  fixed points. Let us study the selection rule for
allowed $n$-point couplings  among twisted states corresponding to $(\theta,
m^{(j)} e)$  for $1\le j\le n$. Their couplings are allowed when
\begin{equation}
 \prod\limits_{j=1}^n (\theta, m^{(j)}\, e)
 ~=~
 (\mathbbm{1},(\mathbbm{1} - \theta)\Lambda)\;.
\end{equation}
The product on the l.h.s.\ evaluates to
\begin{equation}
 \prod\limits_{j=1}^n (\theta, m^{(j)}\, e)
 ~\simeq~
 (\theta^{n},\sum\limits_{j=1}^n m^{(j)}\, e)\;.
\end{equation}
Thus, $n$ localized states with the localization described by
$(\theta,m^{(j)}\,e)$ can only couple if $n$ and $\sum m^{(j)}$ are even. The
latter condition can be understood as a $\mathbbm{Z}_2$ symmetry where the
twisted state $|(\theta, m^{(j)}),\dots \rangle$ (the omission indicates further
quantum numbers) has $\mathbbm{Z}_2$ charge $m^{(j)}$. The $\mathbbm{Z}_2$
transformation can be represented by
\begin{equation}
\sigma_3~=~\left(
\begin{array}{cc}
1 & 0 \\
0 & -1
\end{array}
\right)\; ,
\end{equation}
in a basis where localized states appear as doublets $\left( |(\theta, 0),\dots
\rangle,  |(\theta, e),\dots \rangle \right)$. The requirement that the number
$n$ of involved states be even leads to a second $\mathbbm{Z}_2$ which acts as
$-\mathbbm{1}$ on the above doublets.

Consider, for example, states $x_i$ localized at $m=0$, $y_j$ localized at $m=1$
and bulk fields $b_k$. The symmetries discussed so far restrict allowed
couplings to the form
\begin{equation}\label{eq:example}
 x_{i_1}\cdots x_{i_{n_x}}\,y_{j_1}\cdots y_{j_{n_y}}\,b_{k_1}\cdots b_{k_{n_b}}
 \;,
\end{equation}
with $n_x$ and $n_y$ even. As we shall discuss in the following, one
obtains further relations between the coupling strengths from geometry.

These additional relations hold if the background fields (Wilson lines) vanish.
Then the two fixed points at $m=0,1$ are equivalent. As a consequence, the
Lagrangean is invariant under relabeling $m=0\leftrightarrow 1$. This relabeling
can be interpreted as an $S_2$ permutation of matter fields localized at the
fixed points, and is represented by
\begin{equation}
\sigma_1~=~\left(
\begin{array}{cc}
0 & 1 \\
1 & 0
\end{array}
\right)
\end{equation}
in the basis introduced above. This symmetry relates the coupling strengths,
i.e.\ forces the coupling strength of a term of the structure \eqref{eq:example}
to coincide with an analogous term where $x\leftrightarrow y$. Thus, the flavor
symmetry appearing from $\mathbbm{S}^1/\mathbbm{Z}_2$ is  the multiplicative
closure of $S_2$ and the  $\mathbbm{Z}_2$s, which is denoted by $S_2 \cup
(\mathbbm{Z}_2\times\mathbbm{Z}_2)$. In the case under consideration, the 
subgroup $\mathbbm{Z}_2 \times \mathbbm{Z}_2$, generated by $\sigma_3$ and
$-\mathbbm{1}$, is normal.\footnote{Recall that a subgroup $N$ of a group $G$ is
called normal subgroup if it is  invariant under conjugation; that is, for each
element $n \in N$ and each $g \in G$, the element $g\,n\,g^{-1}$ is still in
$N$. For further details see e.g. \cite{Hall:1976,Lomont:1987,Curtis:1962}.} One
can hence write the flavor symmetry as the semi-direct product $S_2 \ltimes
(\mathbbm{Z}_2 \times \mathbbm{Z}_2)$. The product of the generators of the two
$\mathbbm{Z}_2$s and the $S_2$ leads to the following elements of the
non-Abelian discrete flavor symmetry group:
\begin{equation}
\pm  \mathbbm{1}\;, \quad \pm \sigma_1\;, \quad \pm \I \sigma_2\;,
\quad \pm \sigma_3\;.
\end{equation}
This discrete group is known as $D_4 \equiv S_2 \ltimes (\mathbbm{Z}_2 \times
\mathbbm{Z}_2)$, which is the symmetry of a square.

What we have found so far is that, due to string selection rules and geometry,
superpotential terms enjoy a discrete $D_4$ symmetry where localized states
living at two equivalent fixed points transform as $D_4$ doublets
($\boldsymbol{2}$-plets). Bulk fields are trivial $D_4$ singlets.  Clearly, the
introduction of a (discrete) Wilson line breaks this symmetry explicitly to
$\mathbbm{Z}_2\times\mathbbm{Z}_2$.

In conclusion, one can trade the space group selection rule and invariance under
relabeling  for requiring the Lagrangean to respect a $D_4$ `flavor' symmetry.
It is important to note that the symmetry of the Lagrangean is larger than the
symmetry of internal space. We proceed by applying analogous reasoning to the
remaining building blocks \eqref{eq:BuildingBlocks}.

\subsection{$\boldsymbol{\mathbbm{T}^2/\mathbbm{Z}_3}$ orbifold}

Let us now consider the $\mathbbm{T}^2/\mathbbm{Z}_3$ orbifold which emerges by
dividing the torus $\mathbbm{T}^2_{\mathrm{SU}(3)}$ by its $\mathbbm{Z}_3$
rotational symmetry, i.e.\ the discrete rotation by $120^\circ$. Here
$\mathbbm{T}^2_{\mathrm{SU}(3)}=\mathbbm{R}^2/\Lambda_{\mathrm{SU}(3)}$ with
$\Lambda_{\mathrm{SU}(3)}$ denoting the $\mathrm{SU}(3)$ root lattice  spanned
by two simple roots $e_i$ ($i=1,2$). The $\mathbbm{Z}_3$ twist $\theta$ acts on
the lattice vectors as
\begin{equation}
\theta\, e_1~=~e_2\;, \quad \theta\, e_2~=~-e_1 -e_2\;.
\end{equation}
The sublattice $(1 - \theta) \Lambda$ is spanned by  $e_1 -e_2$ and $3e_1$.
There are three independent fixed points under $\theta$,  which are represented
by 
\begin{equation}
 (\theta, m_1\, e_1)\;,
\end{equation}
with $m_1=0,1,2$. The vector $m_1\, e_1$ is defined up to translations in the
sublattice $(1 - \theta) \Lambda$.

These twisted states residing on the three equivalent fixed points
are degenerate  unless the equivalence of the fixed points is lifted
by the introduction of non-trivial Wilson lines. 
Let us consider an $n$-point coupling of twisted matter  fields
corresponding to $(\theta, m_1^{(j)}\,e_1)$ $(1\le j\le n)$.
According to \eqref{eq:SpaceGroupRule}, this coupling can only be
allowed if the product of space group elements,
\begin{equation}
 \prod\limits_{j=1}^n (\theta,m_1^{(j)}\,e_1)
 ~\simeq~
 (\theta^{n},\sum\limits_{j=1}^n m_1^{(j)}\, e_1)\;,
\end{equation}
is equal to $(\mathbbm{1},(\mathbbm{1}-\theta) \Lambda )$.
That requires
\begin{equation}\label{eq:T2/Z3conditions}
 n~=~3\times (\mathrm{integer})\;,
 \qquad
 \sum\limits_{j=1}^n m_1^{(j)}~=~0 \mod 3\;.
\end{equation}
The first condition is equivalent to demanding that the Lagrangean be invariant
under
\begin{equation}
 \left(\begin{array}{c}
    |(\theta,0),\dots\rangle\\
    |(\theta,e_1),\dots\rangle\\
    |(\theta,2e_1),\dots\rangle
 \end{array}\right)
 ~\to~
 \left(\begin{array}{ccc}
  \omega & 0 & 0\\
  0 & \omega & 0\\
  0 & 0 & \omega
 \end{array}\right)\,
 \left(\begin{array}{c}
    |(\theta,0),\dots\rangle\\
    |(\theta,e_1),\dots\rangle\\
    |(\theta,2e_1),\dots\rangle
 \end{array}\right)\;,
\end{equation}
with $\omega = e^{2 \pi \I/3}$.
The latter condition in \eqref{eq:T2/Z3conditions} corresponds to a
$\mathbbm{Z}_3$ symmetry where the states with $|(\theta,
m_1\,e_1),\dots\rangle$ have the $\mathbbm{Z}_3$ charge $m_1$. That is, one
requires that the Lagrangean be invariant under the $\mathbbm{Z}_3$
transformation
\begin{equation}
 \left(\begin{array}{c}
    |(\theta,0),\dots\rangle\\
    |(\theta,e_1),\dots\rangle\\
    |(\theta,2e_1),\dots\rangle
 \end{array}\right)
 ~\to~
 \left(
 \begin{array}{ccc}
 1 & 0 & 0 \\
 0 & \omega & 0 \\
 0 & 0  &   \omega^2
 \end{array}
 \right)\,
 \left(\begin{array}{c}
    |(\theta,0),\dots\rangle\\
    |(\theta,e_1),\dots\rangle\\
    |(\theta,2e_1),\dots\rangle
 \end{array}\right)\;.
\end{equation}
Furthermore, the effective Lagrangean has an $S_3$ permutation (or relabeling)
symmetry of the  degenerate matter fields living on the three fixed points.
Therefore the combination of selection rules and relabeling symmetry leads to a
discrete flavor symmetry given by the multiplicative closure  $S_3 \cup
(\mathbbm{Z}_3\times\mathbbm{Z}_3)$. As $\mathbbm{Z}_3\times\mathbbm{Z}_3$ is a
normal subgroup, we can write the flavor symmetry group as $S_3 \ltimes
(\mathbbm{Z}_3\times\mathbbm{Z}_3)$. This group has 54 elements and is known as
$\Delta(54)$ in the literature (cf.\ \cite{Fairbairn:1964}).

Matter fields in the $\mathbbm{Z}_3$ orbifold models consist of the untwisted
sector $U$ and $\theta$-twisted sector $T_1$. The $\theta^2$-twisted sector
$T_2$ contains the anti-particles of $T_1$, so that one does not need to treat
it separately.
The untwisted matter fields transform trivially under $\Delta(54)$.
The $T_1$ states transform as $\boldsymbol{3}$-plet
(while the $T_2$ states transform as $\overline{\boldsymbol{3}}$).

In conclusion, the $\mathbbm{T}^2/\mathbbm{Z}_3$ orbifold (or building block)
without Wilson lines enjoys a $\Delta(54)$ flavor symmetry where untwisted and
$T_1$ states transform as singlet and
$\boldsymbol{3}$, respectively.

\subsection{$\boldsymbol{\mathbbm{T}^2/\mathbbm{Z}_4}$  orbifold}
\label{sec:T2/Z4}

\begin{figure}[h]
\centerline{
\subfigure[$\mathbbm{T}^2_\square/\mathbbm{Z}_4$ construction.]{\includegraphics{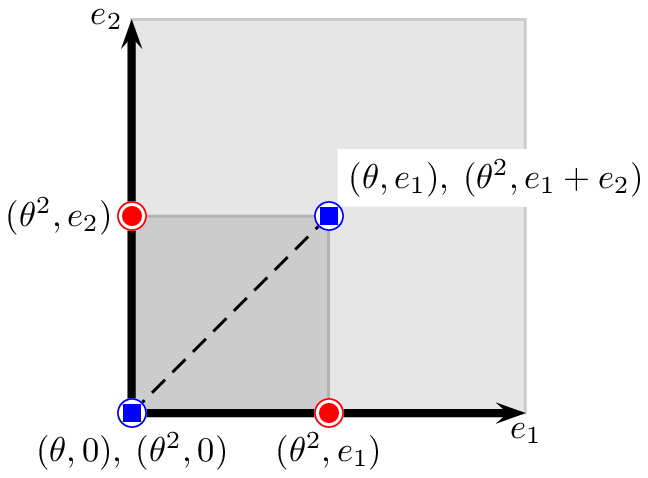}}
\quad
\subfigure[$\mathbbm{T}^2_\square/\mathbbm{Z}_4$ triangle.]{\includegraphics{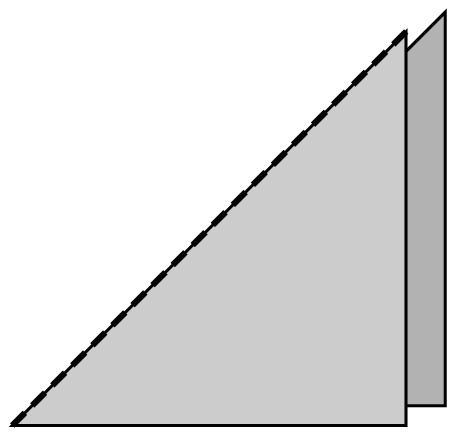}}
}
\caption{(a) $\mathbbm{T}^2_{\mathrm{SO(5)}}$ is defined by two orthonormal
vectors $e_1$ and $e_2$. There are two $\theta$ fixed points, which are
indicated by (blue) squares. In addition one has two $\theta^2$ quasi-fixed
points (red bullets). The fundamental region of the torus consists of the
shaded region, the fundamental region of the orbifold is one quarter thereof.
One can fold the fundamental region along the dashed line and identify adjacent
edges to obtain a triangle with a fore- and a backside (b).}
\label{fig:T2/Z4}
\end{figure}

To construct the $\mathbbm{T}^2/\mathbbm{Z}_4$ orbifold we use the torus
$\mathbbm{T}^2_\square$ which is defined by the two orthonormal torus
translations $e_1$ and $e_2$.\footnote{We could also have used the torus
$\mathbbm{T}_{\mathrm{SO}(5)}$, obtaining the same results.}
The $\mathbbm{Z}_4$ twist acts on $e_1$ and $e_2$ as
\begin{equation}
 \theta\, e_1~=~e_2\;, \qquad \theta\, e_2~=~-e_1\;.
\end{equation}
There are two independent $\mathbbm{Z}_4$ fixed points corresponding to the space group elements
\begin{equation}
 (\theta, m_1\, e_1)\;,
\end{equation}
with $m_1=0,1$ (figure~\ref{fig:T2/Z4}).

The twisted states decompose into $T_1$ and $T_2$ twisted sectors, corresponding
to the space group elements $(\theta,m_1\,e_1)$ and $(\theta^2, \sum_{i =1}^2
m_i\,e_i)$.\footnote{% 
Note that the $T_3$ contains only the anti-particles of
$T_1$, and does therefore not have to be treated separately.} 
Let us first discuss $T_1$. Consider an $n$-point coupling of twisted matter
fields corresponding to $(\theta, m_1^{(j)}e_1)$ $(1\le j\le n)$. This coupling
is allowed only if  the product of space group elements,
\begin{equation}
 \prod\limits_{j=1}^n (\theta,m_1^{(j)}\,e_1)
 ~\simeq~
 (\theta^{n},\sum\limits_{j=1}^n m_1^{(j)}\, e_1)\;,
\end{equation}
is equal to $(\mathbbm{1},\lambda)$ where
$\lambda\in(\mathbbm{1}-\theta)\Lambda$. That requires
\begin{equation}
 n~=~4\times (\mathrm{integer})\;, %\qquad \sum_i m_i~=~0 \mod 2\;.
 \qquad\sum_{j} m_1^{(j)}~=~ 0 \mod 2\;.
\end{equation}
These conditions imply that the Lagrangean is invariant under the two
$\mathbbm{Z}_4$ and $\mathbbm{Z}_2$ transformations
\begin{equation}
 \left(\begin{array}{c}
    |(\theta,0),\dots\rangle\\
    |(\theta,e_1),\dots\rangle
 \end{array}\right)
 ~\to~
 \left(\begin{array}{cc}
  \I & 0 \\
  0 & \I
 \end{array}\right)\,
 \left(\begin{array}{c}
    |(\theta,0),\dots\rangle\\
    |(\theta,e_1),\dots\rangle
 \end{array}\right),
\end{equation}
\begin{equation}
 \left(\begin{array}{c}
    |(\theta,0),\dots\rangle\\
    |(\theta,e_1),\dots\rangle
 \end{array}\right)
 ~\to~
 \left(\begin{array}{cc}
  1 & 0 \\
  0 & -1
 \end{array}\right)\,
 \left(\begin{array}{c}
    |(\theta,0),\dots\rangle\\
    |(\theta,e_1),\dots\rangle
 \end{array}\right),
\end{equation}
respectively.
The former is the $\mathbbm{Z}_4$ transformation generated by  $\I\mathbbm{1}$,
while the latter is the $\mathbbm{Z}_2$ transformation  generated by $\sigma_3$.
Furthermore, the effective Lagrangean has an $S_2$ permutation symmetry
generated by $\sigma_1$.
Thus, the flavor symmetry for $\theta$ twisted states is  the multiplicative
closure of $\{\I\mathbbm{1},\sigma_1,\sigma_3\}$, denoted by $S_2 \cup
(\mathbbm{Z}_4\times\mathbbm{Z}_2)$.
It is quite similar to $D_4$, which is the flavor symmetry on
$\mathbbm{S}^1/\mathbbm{Z}_2$.
The difference is that the above algebra includes
$\mathbbm{Z}_4$ elements $\I\mathbbm{1}$ and $-\I\mathbbm{1}$, and
their products with $D_4$ elements.
Note that the element of the $\mathbbm{Z}_4$ algebra,
$-\mathbbm{1}=(\I)^2\mathbbm{1}$ is included in the $D_4$ algebra. Thus, the
flavor symmetry of $\theta$ twisted matter fields is $(D_4 \times
\mathbbm{Z}_4)/\mathbbm{Z}_2$. The division by $\mathbbm{Z}_2$ implies that we
have to identify the $D_4$ element $-\mathbbm{1}$ with the $Z_4$ element
$(\I)^2\mathbbm{1}$.
This identification has an important meaning for
allowed $\mathbbm{Z}_4$ charges of $D_4$ doublets.
If the flavor symmetry was just $D_4 \times \mathbbm{Z}_4$,
$D_4$ doublets could have an arbitrary $\mathbbm{Z}_4$ charge.
However, since our flavor symmetry is
$(D_4 \times \mathbbm{Z}_4)/\mathbbm{Z}_2$, the
$\mathbbm{Z}_4$ charges of $D_4$ doublets must be equal to
1 or 3, but not 0 or 2 (mod 4).
Hence, the two $T_1$ states correspond to
$\boldsymbol{2}_1$ under $(D_4 \times \mathbbm{Z}_4)/\mathbbm{Z}_2$.
Here, the index denotes the $\mathbbm{Z}_4$ charge.
There is an ambiguity to assign $\mathbbm{Z}_4$ charge 1 or 3 for
the $T_1$ states, but both assignments are equivalent.
With the above assignment, the two $T_3$ states
correspond to $\boldsymbol{2}_3$ under $(D_4 \times \mathbbm{Z}_4)/\mathbbm{Z}_2$.

Let us now study the $T_2$ states. There are four fixed points under the
twist $\theta^2$,
\begin{equation}
(\theta^2,m_1\,e_1 + m_2\,e_2)\;,
\end{equation}
with $m_1, m_2 = 0,1$,
and there are four corresponding $\theta^2$ twisted states,
\begin{equation}
|(\theta^2,m_1\,e_1 + m_2\,e_2),\dots\rangle\; .
\end{equation}
Note that two fixed points $(\theta^2,e_1 )$ and $(\theta^2,e_2 )$ are not fixed
points under the $\theta$ twist, but they are transformed into each other, while the
other two fixed points are fixed points under the $\theta$ twist. Thus, for
states located at  $(\theta^2,e_1 )$ and $(\theta^2,e_2 )$ one has to take
linear combinations to obtain $\theta$ eigenstates as
\cite{Dixon,Kobayashi:1991rp}
\begin{equation}
|(\theta^2,e_1 ),\dots\rangle  \pm |(\theta^2,e_2 ),\dots\rangle\; .
\end{equation}
These four space group elements, $(\theta^2,m_1e_1 + m_2e_2)$, can be obtained
as products of two space group elements  $(\theta, m_1^{(1)}e_1)$ and $(\theta,
m_1^{(2)}e_1)$ according to
\begin{align}
 (\theta^2,0)&=~(\theta,0)(\theta,0)\;, &\qquad
(\theta^2,e_1) &=~(\theta,e_1)(\theta,0)\;,  \nonumber\\
 (\theta^2,e_2) &=~(\theta,0)(\theta,e_1)\;, &\qquad
(\theta^2,e_1+e_2) &=~(\theta,e_1)(\theta,e_1)\;.
\end{align}
Since the selection rules for allowed couplings  including both $\theta$ twisted
states and  $\theta^2$ twisted states are controlled by the space group, the
above products determine how  four $\theta^2$ twisted states transform under
$(D_4 \times \mathbbm{Z}_4)/\mathbbm{Z}_2$.
The two $\theta$ twisted states  $|(\theta,m_1\,e_1 )),\dots\rangle $ with
$m_1=0,1$ transform as $\boldsymbol{2}_1$  under  $(D_4 \times
\mathbbm{Z}_4)/\mathbbm{Z}_2$,  and the product of two such doublets yields
\begin{equation}\label{eq:mult}
 \boldsymbol{2}_1 \times \boldsymbol{2}_1
 ~=~(\boldsymbol{1}_{A_1})_2 + (\boldsymbol{1}_{B_1})_2
 + (\boldsymbol{1}_{A_2})_2 + (\boldsymbol{1}_{B_2})_2\;,
\end{equation}
where $\boldsymbol{1}_{A_1,B_1,A_2,B_2}$ denote the four types of $D_4$ singlets
(see Appendix \ref{app:D4}).
Thus, four $\theta^2$ twisted states can be expressed in terms of four types of
$D_4$ singlets,
\begin{eqnarray}
 (\boldsymbol{1}_{A_1})_2 &:&
 |(\theta^2,0 ),\dots\rangle  + |(\theta^2,e_1+e_2 ),\dots\rangle\; ,
 \nonumber\\
 (\boldsymbol{1}_{B_2})_2 &:&
 |(\theta^2,0 ),\dots\rangle  - |(\theta^2,e_1+e_2 ),\dots\rangle\; ,
 \nonumber\\
 (\boldsymbol{1}_{B_1})_2 &:&
 |(\theta^2,e_1 ),\dots\rangle  + |(\theta^2,e_2 ),\dots\rangle\; ,
 \nonumber\\
 (\boldsymbol{1}_{A_2})_2 &:&
 |(\theta^2,e_1 ),\dots\rangle  - |(\theta^2,e_2 ),\dots\rangle \;.
\end{eqnarray}

If we consider couplings including only $\theta^2$ twisted states and untwisted
states, the flavor symmetry of $\theta^2$ twisted states is the same as the
discrete symmetry for $T_1$ states on $\mathbbm{T}^2/\mathbbm{Z}_2$, i.e.\ $(D_4
\times D_4)/\mathbbm{Z}_2$ (we will discuss this further in
section~\ref{sec:symmetry2}). However,  couplings including both $\theta$ and
$\theta^2$  twisted states enjoy only $(D_4 \times
\mathbbm{Z}_4)/\mathbbm{Z}_2$.

As an example, let us consider the allowed 3-point couplings, $T_1 T_1 T_2$. To
obtain them, we decompose the product of two $T_1$ doublets
\begin{equation}
 |\boldsymbol{2}_1,1\rangle~=~\left(\begin{array}{c}
    |(\theta,0),1\rangle\\ |(\theta,e_1),1\rangle
 \end{array}\right)
 \quad\text{and}\quad
 |\boldsymbol{2}_1,2\rangle~=~\left(\begin{array}{c}
    |(\theta,0),2\rangle\\ |(\theta,e_1),2\rangle
 \end{array}\right)
\end{equation}
according to \eqref{eq:mult} into four singlets,
\begin{eqnarray}
 \boldsymbol{1}_{A_1} & : &
    |(\theta,0),1\rangle\,|(\theta,0),2\rangle
    +
    |(\theta,e_1),1\rangle\,|(\theta,e_1),2\rangle
 \;,\nonumber\\
 \boldsymbol{1}_{B_2} & : &
    |(\theta,0),1\rangle\,|(\theta,0),2\rangle
    -
    |(\theta,e_1),1\rangle\,|(\theta,e_1),2\rangle
 \;,\nonumber\\
 \boldsymbol{1}_{B_1} & : &
    |(\theta,0),1\rangle\,|(\theta,e_1),2\rangle
    +
    |(\theta,e_1),1\rangle\,|(\theta,0),2\rangle
 \;,\nonumber\\
 \boldsymbol{1}_{A_2} & : &
    |(\theta,0),1\rangle\,|(\theta,e_1),2\rangle
    -
    |(\theta,e_1),1\rangle\,|(\theta,0),2\rangle
 \;.
\end{eqnarray}
Now seek for invariant product with $T_2$ states. Only the following products
are invariant:
\begin{equation}
 A_1 \ A_1, \ A_2 \ A_2, \ B_1 \ B_1, \ B_2 \ B_2\;.
\end{equation}
From $A_1^2$ and $B_2^2$ we find
the following allowed superpotential terms (we are using the states as synonyms for the superfields)
\begin{eqnarray}
&a \ [|(\theta,0),1\rangle\, |(\theta,0),2\rangle
 + |(\theta,e_1),1\rangle\, |(\theta,e_1),2\rangle] \
 [|(\theta^2,0)\rangle + |(\theta^2,e_1+e_2)\rangle]\;, &  \nonumber \\
&b \ [|(\theta,0),1\rangle\, |(\theta,0),2\rangle -
|(\theta,e_1),1\rangle\, |(\theta,e_1),2\rangle]
\ [|(\theta^2,0)\rangle - |(\theta^2,e_1+e_2)\rangle]\;. &
 \end{eqnarray}
We can re-arrange the couplings above and obtain for the couplings of physical
states
\begin{eqnarray}
 W& \supset & c_1 \, [|(\theta,0),1\rangle\, |(\theta,0),2\rangle \,
    |(\theta^2,0)\rangle
    + |(\theta,e_1),1\rangle\,|(\theta,e_1),2\rangle \
    |(\theta^2,e_1+e_2)\rangle]
 \nonumber\\
 & & {}+c_2  \ [|(\theta,0),1\rangle\, |(\theta,0),2\rangle \,
 |(\theta^2,e_1 + e_2)\rangle +
 |(\theta,e_1),1\rangle\, |(\theta,e_1),2\rangle \, |(\theta^2,0)\rangle]
\end{eqnarray}
with $c_1 = a + b$, $c_2 = a - b$. From $B_1^2$ and $A_2^2$ we get
\begin{eqnarray}
 W &\supset & c\, \left[
    |(\theta,0),1\rangle\,|(\theta,e_1),2\rangle
    +
    |(\theta,e_1),1\rangle\,|(\theta,0),2\rangle\right]\,
  [|(\theta^2,e_1)\rangle + |(\theta^2,e_2)\rangle]
 \nonumber\\
 & & {}+d\, \left[
    |(\theta,0),1\rangle\,|(\theta,e_1),2\rangle
    -
    |(\theta,e_1),1\rangle\,|(\theta,0),2\rangle
 \right]\,
  [|(\theta^2,e_1)\rangle - |(\theta^2,e_2)\rangle] \;.
  \label{eq:cd}
\end{eqnarray}
The re-arrangement of couplings then analogously reads
\begin{eqnarray}
 W &\supset & c_3\, \left[
    |(\theta,0),1\rangle\,|(\theta,e_1),2\rangle\,
  |(\theta^2,e_1)\rangle + |(\theta,e_1),1\rangle\,|(\theta,0),2\rangle|(\theta^2,e_2)\rangle\right]
 \nonumber\\
 & & {}+c_4\, \left[
    |(\theta,0),1\rangle\,|(\theta,e_1),2\rangle\,
    |(\theta^2,e_2)\rangle + |(\theta,e_1),1\rangle\,|(\theta,0),2\rangle\,|(\theta^2,e_1)\rangle\right] 
\end{eqnarray}
with $c_3 = c + d$, $c_4 = c - d$. The higher-dimensional allowed couplings can
be obtained analogously. From the geometry of the setup one infers that
$c_3=c_4$ such that $d=0$ in \eqref{eq:cd}.\footnote{We thank P.~Vaudrevange for
pointing this out to us.} The vanishing of the $d$-term can also be inferred
from gauge invariance. This will be discussed in detail elsewhere.

\subsection{$\boldsymbol{\mathbbm{T}^2/\mathbbm{Z}_6}$ orbifold}

The 2-dimensional $\mathbbm{Z}_6$ orbifold is obtained as
$\mathbbm{T}^2_{\mathrm{SU}(3)}/\mathbbm{Z}_6$.\footnote{We could have
considered the $\mathrm{G}_2$ lattice, obtaining the same result.}
% $\mathbbm{R}^2/(\Lambda_{\mathrm{SU}(3)}\times \mathbbm{Z}_6)$.
The $\mathbbm{Z}_6$ twist is defined for the $\mathrm{SU}(3)$ simple roots $e_1$
and $e_2$ as
\begin{equation}
\theta\, e_1~=~e_1+ e_2\;, \qquad \theta\,( e_1+ e_2)~=~e_2\;, \qquad
\theta\, e_2~=~-e_1\;.
\end{equation}
The sublattice $(1 - \theta) \Lambda_{\mathrm{SU}(3)}$ is the same as
$\Lambda_{\mathrm{SU}(3)}$.
That implies that there is a single (independent) fixed point
under the $\mathbbm{Z}_6$ twist $\theta$.
That is, this orbifold model does not include a
non-Abelian flavor structure.

Notice that couplings involving higher twisted sectors only do enjoy non-Abelian
discrete symmetries. Analogously to the $\mathbbm{Z}_4$ case, these symmetries disappear
as soon as $T_1$ states enter the couplings.

\subsection{$\boldsymbol{\mathbbm{T}^4/\mathbbm{Z}_8}$ orbifold}

The 4-dimensional $\mathbbm{Z}_8$ orbifold can be obtained as
$\mathbbm{T}^4_{\mathrm{SO}(9)}/\mathbbm{Z}_8$, where
$\mathbbm{T}^4_{\mathrm{SO}(9)}$ is the 4-torus based on the SO(9) Lie lattice,
which is spanned by the basis vectors $e_i$ $(i=1,2,3,4)$. The $\mathbbm{Z}_8$
twist transforms the latter
\begin{equation}
\theta\, e_1~=~e_{2}\;, \qquad
\theta\, e_2~=~e_{3}\;, \qquad\theta\, e_3~=~\sum_{j=1}^3e_j + 2e_4\;,
\qquad \theta\, e_4~=~-\sum_{j=1}^4e_j\;.
\end{equation}
The sublattice $(1 - \theta)\Lambda$ is spanned by
$e_1, e_2, e_3$ and $2e_4$.
Thus, there are two fixed points under $\theta$,
\begin{equation}
(\theta,m_4\,e_4),
\end{equation}
for $m_4=0,1$.

The flavor symmetry of $T_1$ on $\mathbbm{T}^4/\mathbbm{Z}_8$
is quite similar to the flavor symmetry of $T_1$ on
$\mathbbm{T}^2/\mathbbm{Z}_4$.
The former is the closure algebra,
$S_2 \cup (\mathbbm{Z}_8\times\mathbbm{Z}_2)$,
where $\mathbbm{Z}_8$ transforms as
\begin{equation}
 \left(\begin{array}{c}
        |(\theta,0),\dots\rangle\\
        |(\theta,e_4),\dots\rangle
 \end{array}\right)
 ~\to~
 \left(\begin{array}{cc}
  \rho & 0 \\
  0 & \rho
 \end{array}\right)\,
 \left(\begin{array}{c}
        |(\theta,0),\dots\rangle\\
        |(\theta,e_4),\dots\rangle
 \end{array}\right),
\end{equation}
with $\rho = e^{\I\pi/4}$.
In addition, $S_2$ and $\mathbbm{Z}_2$ transformations are represented
in the above basis by $\sigma_1$ and $\sigma_3$, respectively.
Thus, the flavor symmetry is written as
$S_2 \cup (\mathbbm{Z}_8\times\mathbbm{Z}_2) =
(D_4\times \mathbbm{Z}_8)/\mathbbm{Z}_2$, where the
division by $\mathbbm{Z}_2$ implies that we identify
the $D_4$ element $-\mathbbm{1}$ with the $\mathbbm{Z}_8$
element $\rho^4 \mathbbm{1}$.
Therefore, the two $T_1$ states correspond to
${\bf 2}_1$ under $(D_4\times \mathbbm{Z}_8)/\mathbbm{Z}_2$,
where the index denotes $\mathbbm{Z}_8$ charge.

In addition, the $\theta^2$-twisted sector
$T_2$ has four independent fixed points,
\begin{equation}
(\theta^2,m_3\,e_3+m_4\,e_4),
\end{equation}
with $m_1,m_2=0,1$.
Through a discussion similar to section~\ref{sec:T2/Z4}, we find that four $T_2$
states on the above fixed points correspond to four types of singlets,
\begin{equation}
(\boldsymbol{1}_{A_1})_2 + (\boldsymbol{1}_{B_1})_2
+ (\boldsymbol{1}_{A_2})_2 + (\boldsymbol{1}_{B_2})_2,
\end{equation}
under $(D_4\times \mathbbm{Z}_8)/\mathbbm{Z}_2$.

The $\theta^3$-twisted sector $T_3$ has the same structure
of fixed points as the $\theta$-twisted sector.
Thus, the two $T_3$ states correspond to
$\boldsymbol{2}_3$ under $(D_4\times \mathbbm{Z}_8)/\mathbbm{Z}_2$.

Moreover, the $\theta^4$-twisted sector $T_4$ has 16 independent
fixed points,
\begin{equation}
 (\theta^4,\sum_{i=1}^4 m_i\,e_i),
\end{equation}
with $m_i=0,1$.
The corresponding 16 $T_4$ states must be $D_4$ singlets with
the $\mathbbm{Z}_8$ charge 4.
From the $D_4$ multiplication law
\begin{equation}
 (\boldsymbol{1}_{A_1} + \boldsymbol{1}_{B_1} +\boldsymbol{1}_{B_2} +
\boldsymbol{1}_{A_2})
\times
(\boldsymbol{1}_{A_1} + \boldsymbol{1}_{B_1} +\boldsymbol{1}_{B_2} +
\boldsymbol{1}_{A_2})
~=~
4 \times (\boldsymbol{1}_{A_1} + \boldsymbol{1}_{B_1} +\boldsymbol{1}_{B_2}
+ \boldsymbol{1}_{A_2})
\end{equation}
it follows that the 16 $T_4$ states correspond to
\begin{equation}
4 \times \left( (\boldsymbol{1}_{A_1})_4 + (\boldsymbol{1}_{B_1})_4
  +(\boldsymbol{1}_{B_2})_4 + (\boldsymbol{1}_{A_2})_4 \right)\;.
\end{equation}
If we consider couplings involving only $T_2$ or $T_4$ states, the flavor
symmetry would be larger, but it gets broken if $T_1$ states enter.

\subsection{$\boldsymbol{\mathbbm{T}^4/\mathbbm{Z}_{12}}$ orbifold}

There is only one fixed point on
the $\mathbbm{T}^4/\mathbbm{Z}_{12}$ orbifold.
Thus, the situation is the same as in $\mathbbm{T}^2/\mathbbm{Z}_6$,
i.e.\ there is no non-Abelian flavor symmetry.

\subsection{$\boldsymbol{\mathbbm{T}^6/\mathbbm{Z}_7}$ orbifold}

For completeness let us consider the
$\mathbbm{T}^6_{\mathrm{SU}(7)}/\mathbbm{Z}_7$ orbifold. It is obtained as
$\mathbbm{R}^6/(\Lambda_{\mathrm{SU}(7)} \times \mathbbm{Z}_7)$, where
$\Lambda_{\mathrm{SU}(7)}$ denotes the $\mathrm{SU}(7)$ root lattice spanned by
six simple roots, $e_i$ $(1\le i\le 6)$. The $\mathbbm{Z}_7$ twist transforms
these roots as
\begin{equation}
\theta e_i~=~e_{i+1}\;, \qquad e_6~=~-\sum_{j=1}^6 e_j\;,
\end{equation}
for $1\le i\le 5$. The sublattice $(1 -\theta)\, \Lambda$ is spanned by $e_i -
e_{i+1}$ and $7e_1$, and there are seven independent fixed points under
$\theta$,
\begin{equation}
(\theta, m\,e_1)\;,
\end{equation}
with $0\le m\le 6$.
The other $T_k$ sectors with $k \neq 0$ have the same fixed point structure.

The flavor symmetry is obtained in a way similar to the extension of the
$\mathbbm{Z}_3$ orbifold. That is, the states at these seven fixed points have $\mathbbm{Z}_7$
charges, and the effective Lagrangean has a permutation symmetry of $S_7$. Therefore, the flavor symmetry of the $\mathbbm{Z}_7$ orbifold
is the multiplicative closure of $S_7$ and the $\mathbbm{Z}_7$. To determine the dimension of this group, it is
useful to rewrite it as a semi-direct product. It is easy
to see that the diagonal matrices corresponding to $\mathbbm{Z}_7$
transformations are obtained as products of the following six matrices:
\[\begin{array}{lll}
\diag(1,\rho,\rho^2,\rho^3,\rho^4,\rho^5,\rho^6)\;,
& \diag(\rho,\rho,\rho,\rho,\rho,\rho,\rho)\;,
& \diag(\rho^1,\rho^6,\rho^2,\rho^3,\rho^4,\rho^5,1)\;, \\
\diag(1,\rho,\rho^6,\rho^3,\rho^4,\rho^5,\rho^2) \;,
& \diag(1,\rho^2,\rho,\rho^6,\rho^4,\rho^5,\rho^3)\;,
& \diag(1,\rho^3,\rho^2,\rho,\rho^6,\rho^5,\rho^4)\;,
\end{array}
\]
where $\rho=e^{2 \pi \I/7}$.
With these generators , the flavor symmetry can be expressed as $S_7 \ltimes
(\mathbbm{Z}_7)^6$. Its order is equal to $7!\cdot 7^6$, and it is quite large.
The $T_k$ matter fields correspond to septets, and $T_{7-k}$ fields correspond
to their conjugates.

\section{Flavor symmetries of `factorizable' orbifolds}
\label{sec:CombiningBuildingBlocks}

If the torus $\mathbbm{T}^d$ of an orbifold factorizes, the orbifold
is often called `factorizable' in the literature although it cannot
be regarded as a direct product of lower-dimensional orbifolds.  The
symmetries of such orbifolds can be obtained by combining flavor
symmetries of the building blocks. However, the  resulting flavor symmetries are generally not
direct products of the symmetries of the building blocks, since
orbifolds do not really factorize. We discuss the subtleties in the case of
$\mathbbm{T}^2/\mathbbm{Z}_2$, and give an overview of how to obtain
the flavor symmetries of `products' of building blocks in other
higher-dimensional orbifolds.

\subsection{$\boldsymbol{\mathbbm{T}^2/\mathbbm{Z}_2}$ as a `product' of two
$\boldsymbol{\mathbbm{S}^1/\mathbbm{Z}_2}$ orbifolds}
\label{sec:symmetry2}

\subsubsection{Generic situation}

\begin{figure}[h]
\centerline{\includegraphics{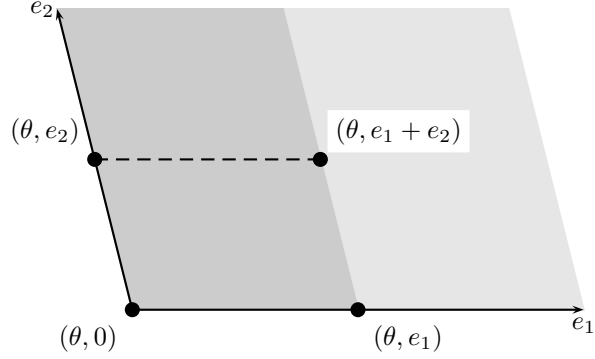}}
\caption{$\mathbbm{T}^2/\mathbbm{Z}_2$. Points which are related by a reflection
at the origin are identified. The fundamental region of the orbifold (dark
gray region) is half of the fundamental region of the torus (gray region). By
folding it along the dashed line and identifying the edges one obtains a
`ravioli' (or `pillow') with the fixed points being the corners.}
\label{fig:T2overZ2}
\end{figure}

The $\mathbbm{T}^2/\mathbbm{Z}_2$ orbifold is obtained by dividing the torus by
the reflection at the origin.
% obtained as a direct product of
% two 1D orbifolds, $\mathbbm{S}^1/\mathbbm{Z}_2$.
The 2D torus is defined by a 2D lattice which is  spanned by $e_i$ $(i=1,2)$.
The $\mathbbm{Z}_2$ twist $\theta$ acts then on the $e_i$ as
\begin{equation}
\theta\, e_i~=~- e_i\;.
\end{equation}
There are four fixed fixed points $(\theta,m_1\,e_1+m_2\,e_2)$ where $m_i=0,1$
(figure~\ref{fig:T2overZ2}).

From the space group rule \eqref{eq:SpaceGroupRule} one infers that a coupling
involving $n$ localized states
$|(\theta,m^{(j)}_1e_1+m^{(j)}_2e_2),\dots\rangle$ can be allowed only if
\begin{eqnarray}
 n & \text{is} & \text{even}\;,\nonumber\\
 \sum_{j=1}^n m_i^{(j)} & \text{is} & \text{even}\;,\quad i=1,2\;.
 \label{eq:T2overZ2}
\end{eqnarray}
As before, the selection rules can be rewritten in a different way. In a basis where the localized
states appear as $\boldsymbol{4}$-plets,
$|\boldsymbol{4}\rangle=\left(|(\theta,0),\dots\rangle,|(\theta,e_1),\dots\rangle,
|(\theta,e_2),\dots\rangle,|(\theta,e_1+e_2),\dots\rangle \right)$, the
selection rules \eqref{eq:T2overZ2} allow couplings only if they are invariant
under $|\boldsymbol{4}\rangle\to A\,|\boldsymbol{4}\rangle$ with
\begin{equation}
 A~\in~\{P,Q,R\}~=~\left\{
 \left(\begin{array}{cc}
    \sigma_3 & 0 \\
    0 & \sigma_3
 \end{array}\right),\,
 \left(\begin{array}{cc}
    \mathbbm{1}_2 & 0 \\
    0 & -\mathbbm{1}_2
 \end{array}\right),\,
 \left(\begin{array}{cc}
    -\mathbbm{1}_2 & 0 \\
    0 & -\mathbbm{1}_2
 \end{array}\right) \right\}\;.
\end{equation}

Again, in the absence of Wilson lines, the fixed points are equivalent. Hence
the Lagrangean is invariant under relabeling
\begin{equation}
m_i~\to~m_i+1 \mod2\;,\quad i=1,2\;.
\end{equation}
This relabeling corresponds, as before, to a permutation symmetry.
Here, we have two separate permutations that can be represented as
$|\boldsymbol{4}\rangle\to S\,|\boldsymbol{4}\rangle$ and
$|\boldsymbol{4}\rangle\to S'\,|\boldsymbol{4}\rangle$ where
\begin{equation}\label{eq:P}
 S~=~
  \left(\begin{array}{cc}
    \sigma_1 & 0 \\
    0 & \sigma_1
 \end{array}\right)\;,
 \quad
 S'~=~
  \left(\begin{array}{cc}
    0 & \mathbbm{1}_2 \\
    \mathbbm{1}_2 & 0
 \end{array}\right)
\end{equation}
in the above basis.

The connection between the generators $\{\sigma_1,\, \sigma_3\}$ of the flavor
symmetry for $\mathbbm{S}^1/\mathbbm{Z}_2$ and that for
$\mathbbm{T}^2/\mathbbm{Z}_2$ can be seen from
\begin{align}
 Q & =~
 \sigma_3 \otimes \mathbbm{1}_{2 \times 2}\;,
 &
 S'& =~ \sigma_1 \otimes \mathbbm{1}_{2 \times 2}\;,\nonumber\\
 P & =~ \mathbbm{1}_{2 \times 2} \otimes \sigma_3\;,
 &
 S & =~ \mathbbm{1}_{2 \times 2} \otimes \sigma_1 \; ,
\end{align}
where $\otimes$ denotes the Kronecker product. This construction can easily be
generalized to other higher-dimensional orbifolds.

The non-Abelian symmetry group arising from
$\mathbbm{T}^2/\mathbbm{Z}_2$ is hence comprised of the
multiplicative closure of the above matrices $P$, $Q$, $R$, $S$ and
$S'$. We obtain a flavor symmetry $(S_2 \times
S_2)\ltimes(\mathbbm{Z}_2\times\mathbbm{Z}_2\times\mathbbm{Z}_2)$.
This symmetry group has 32 elements, and is a subgroup of $D_4\times
D_4$ which has 64 elements. It is very similar to the Dirac group
(see, e.g., \cite{Lomont:1987}). The reason for having less symmetry
than what one would have for the product space
$(\mathbbm{S}^1/\mathbbm{Z}_2)\times
(\mathbbm{S}^1/\mathbbm{Z}_2)$ (i.e.\ $D_4 \times D_4$)  is
that in $\mathbbm{T}^2/\mathbbm{Z}_2$ the automorphism $\theta$
reflects both $e_i$ simultaneously. Therefore one has a
$\mathbbm{Z}_2$ less, and correspondingly half as many elements, as
it should be. Because both $D_4$ factors have a common (`diagonal')
$\mathbbm{Z}_2$, we call the flavor symmetry group $(D_4\times
D_4)/\mathbbm{Z}_2$. Note that the `would-be'
$(\boldsymbol{2},\boldsymbol{2})$ under $D_4\times D_4$ transforms
as an irreducible $\boldsymbol{4}$-dimensional representation under
$(D_4\times D_4)/\mathbbm{Z}_2$.

\subsubsection{Symmetry enhancement}

\begin{figure}[h]
\centerline{\subfigure[$\mathbbm{T}^2_{\mathrm{SU}(3)}/\mathbbm{Z}_2\,.$]{%
    \includegraphics{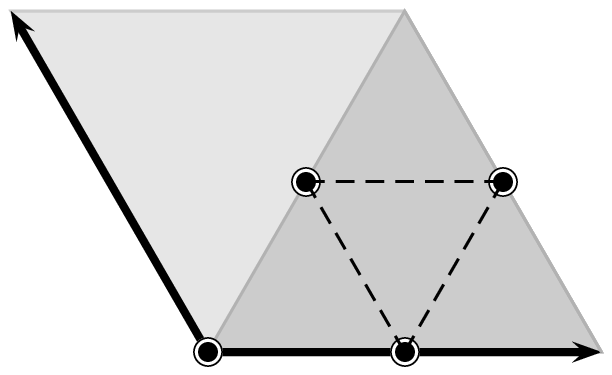}}
\qquad
\subfigure[Tetrahedron.]{\includegraphics[scale=0.7]{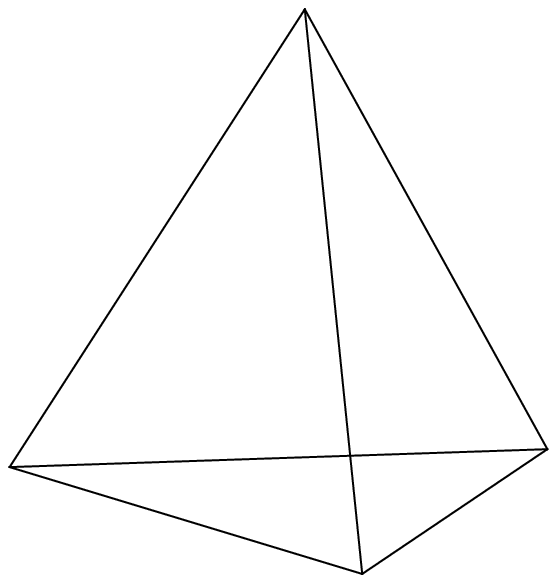}}}
\caption{If the $\mathbbm{T}^2$ lattice vectors have equal length and enclose
$120^\circ$, one can also fold the fundamental region of
$\mathbbm{T}^2/\mathbbm{Z}_2$ to a tetrahedron.}
\label{fig:tetrahedron}
\end{figure}

An interesting situation arises when the torus $\mathbbm{T}^2$ has
special symmetries. Consider the case where $e_1$ and $e_2$ have the
same length, and enclose an angle of $120^\circ$. Then the symmetry
gets enhanced since the distances between all orbifold fixed points
coincide. One may now envisage the orbifold as a regular tetrahedron
(figure~\ref{fig:tetrahedron}) with the corners corresponding to the
fixed points \cite{Dixon:1986qv} (this observation has been recently revisited \cite{Altarelli:2006kg}).

Clearly, the tetrahedron is invariant under a discrete rotation by $120^\circ$
about an axis that goes through one corner and hits the opposite surface
orthogonally. This operation is represented by
\begin{eqnarray}
T~= ~\left( \begin{array}{cccc} 1 & 0 & 0 & 0 \\
0 & 0 & 1 & 0 \\
0 & 0 & 0 & 1 \\
0 & 1 & 0 & 0
 \end{array}
\right)\;,\quad T\,S\;,\quad T\,S'\;,\quad T\,S\,S'\;.
\end{eqnarray}
The full relabeling symmetry includes the symmetry of the tetrahedron, i.e.\
$A_4$. $A_4$ arises as multiplicative closure of the $\mathbbm{Z}_2$ and
$\mathbbm{Z}_3$ groups with elements $\{\mathbbm{1},S\}$ and
$\{\mathbbm{1},T,T^2\}$, respectively.

However, $A_4$ is not the full relabeling symmetry because the
geometric relations between the fixed points do not change upon
reflections (which, however, change the orientation, and are
therefore not symmetries of the solid). The full relabeling symmetry
is therefore $S_4$.
As before, the flavor symmetry group is to be amended by the symmetries arising
from the space group rules, i.e.\ one has to include the elements $P$, $Q$ and
$R$.
The flavor symmetry gets therefore enhanced to
$S_4\ltimes(\mathbbm{Z}_2\times\mathbbm{Z}_2\times\mathbbm{Z}_2)$, which is
known as SW$_4$ in the literature \cite{Baake:1981qe}, and has $4!\cdot 2^3=192$
elements.\footnote{We would like to thank C.~Hagedorn for making us aware of
reference~\cite{Baake:1981qe} and for pointing out its relevance for our
investigations.}  As before, the symmetry of the Lagrangean is larger than the
symmetry of internal space.
Recently an $A_4$ subgroup of the
tetrahedral compactification symmetry SW$_4$ has been considered
in the framework of neutrino mixing matrices in \cite{Altarelli:2006kg}.

Another interesting case with enhanced symmetries is if (i) $e_1$ and $e_2$
enclose an angle of $90^\circ$ and (ii) $e_1$ and $e_2$ have equal length. The
orbifold can then be envisaged as a perfect square with a fore- and a backside.
One now has a relabeling symmetry consisting of cyclic permutations
($\mathbbm{Z}_4$), generated by
\begin{equation}
 S''~=~\left(\begin{array}{cccc}
  0 & 1 & 0 & 0 \\
  0 & 0 & 1 & 0 \\
  0 & 0 & 0 & 1 \\
  1 & 0 & 0 & 0
  \end{array}\right)\;,
\end{equation}
plus the flips generated by $S$ and $S'$. The multiplicative closure
of the operations represented by $P$, $Q$, $R$, $S$, $S'$ and $S''$
is $D_4\ltimes (\mathbbm{Z}_2)^3$, and has 64 elements. It has 16 conjugacy classes,
two four-dimensional, six two-dimensional and eight one-dimensional
irreducible representations. From the construction, it is obvious
that this order 64 group is a subgroup of the above-mentioned
SW$_4$. Clearly, it contains the order 32 symmetry of the generic
$\mathbbm{T}^2/\mathbbm{Z}_2$ as a subgroup.

An important remark, applicable to both cases above,
concerns the symmetry breakdown occurring when the angle between
$e_1$ and $e_2$ and/or their length ratio changes. Both the angle
and the ratio are parametrized by a field $Z$, called complex
structure modulus in the literature. Hence symmetry breakdown can be
described by a departure of the vacuum expectation value (VEV) of
$Z$ from its symmetric value. That is, the couplings between
localized states are $Z$-dependent, and the coupling strengths
respect an enhanced symmetry if $Z$ takes special values.

\subsection{Other combinations of building blocks}

From the above discussion it is clear how to obtain the flavor symmetries of other
combinations of building blocks. In general, these emerge as products of the flavor
symmetries of the building blocks with a common (`diagonal') $\mathbbm{Z}_n$
sub-group identified. As an example consider $\mathbbm{T}^4/\mathbbm{Z}_2$ where
the flavor symmetry is $(D_4\times D_4\times D_4\times D_4)/(\mathbbm{Z}_2)^3=
(S_2\times S_2\times S_2\times S_2)\ltimes (\mathbbm{Z}_2^5)$.

Exceptions to this statement occur if independent twists act on the building
blocks. An example for such a situation is the
$\mathbbm{Z}_6-\mathrm{II}=\mathbbm{Z}_3\times\mathbbm{Z}_2$ orbifold which
enjoys a $\Delta(54)\times(D_4\times D_4)/\mathbbm{Z}_2$ symmetry coming from
the building blocks $\mathbbm{T}^2/\mathbbm{Z}_3$ and
$\mathbbm{T}^2/\mathbbm{Z}_2$.

We note that such large flavor symmetry groups would not be expected in
realistic models because they only arise in the absence of Wilson
lines, where one often obtains too many families. Moreover, Wilson
lines are generically needed in order to reduce the gauge symmetry
to the standard model gauge group (amended by a `hidden sector').

\begin{table}[!h]
\begin{center}
\begin{tabular}{|c|c|c|c|}  \hline
orbifold & flavor symmetry & twisted sector &  string fundamental
states
\\ \hline
$\mathbbm{S}^1/\mathbbm{Z}_2$ & $D_4$ $=S_2\ltimes (\mathbbm{Z}_2 \times \mathbbm{Z}_2)$ & untwisted sector & $\boldsymbol{1}$ \\
   &  & $\theta$-twisted sector & {\bf  2} \\
\hline $\mathbbm{T}^2/\mathbbm{Z}_2$ & $(D_4\times
D_4)/\mathbbm{Z}_2$  $=(S_2\times S_2)\ltimes \mathbbm{Z}_2^3$  &
untwisted
sector & $\boldsymbol{1}$ \\
   &   & $\theta$-twisted sector &
   $\boldsymbol{4}$ \\  \hline
$\mathbbm{T}^2/\mathbbm{Z}_3$ &  & untwisted sector &
$\boldsymbol{1}$ \\
   & $\Delta(54)$ $=S_3\ltimes (\mathbbm{Z}_3 \times \mathbbm{Z}_3)$ & $\theta$-twisted sector & {\bf 3} \\
   &  & $\theta^2$-twisted sector & ${\bar {\bf 3}}$ \\ \hline
$\mathbbm{T}^2/\mathbbm{Z}_4$ &   & untwisted sector &
$\boldsymbol{1}$ \\
   & $(D_4 \times \mathbbm{Z}_4)/\mathbbm{Z}_2$  & $\theta$-twisted sector & {\bf 2} \\
   & & $\theta^2$-twisted sector & $\boldsymbol{1}_{A_1} + \boldsymbol{1}_{B_1} +\boldsymbol{1}_{B_2} + \boldsymbol{1}_{A_2}$ \\
\hline
$\mathbbm{T}^2/\mathbbm{Z}_{6}$ & trivial & & \\ \hline
$\mathbbm{T}^4/\mathbbm{Z}_8$ &  & untwisted sector &
$\boldsymbol{1}$ \\
   &    & $\theta$-twisted sector & {\bf 2} \\
   &  $(D_4 \times \mathbbm{Z}_8) / \mathbbm{Z}_2$  & $\theta^2$-twisted sector &
$ \boldsymbol{1}_{A_1} + \boldsymbol{1}_{B_1} +\boldsymbol{1}_{B_2} + \boldsymbol{1}_{A_2}$ \\
   & & $\theta^3$-twisted sector & {\bf 2} \\
   & & $\theta^4$-twisted sector &
$4 \times (\boldsymbol{1}_{A_1} + \boldsymbol{1}_{B_1}
+\boldsymbol{1}_{B_2} + \boldsymbol{1}_{A_2})$
\\  \hline
$\mathbbm{T}^4/\mathbbm{Z}_{12}$ & trivial & & \\ \hline
$\mathbbm{T}^6/\mathbbm{Z}_7$ & & untwisted sector & $\boldsymbol{1}$ \\
    & $S_7 \ltimes (\mathbbm{Z}_7)^6$ & $\theta^k$-twisted sector &
    $\boldsymbol{7}$ \\
   & & $\theta^{7-k}$-twisted sector & ${\bar {\bf 7}}$ \\
\hline
\end{tabular}
\end{center}
\caption{Non-Abelian discrete flavor symmetries of the building
blocks.}
\label{tab-3}
\end{table}

It is also clear that symmetry enhancement occurs in various orbifolds.
In table \ref{tab-0} (a), there are three more cases where the
flavor symmetry can be enlarged: $\mathbbm{T}^4 /\mathbbm{Z}_4$,
$\mathbbm{T}^4 /\mathbbm{Z}_3$, $\mathbbm{T}^6 / \mathbbm{Z}_3$. A
similar analysis as above shows that for the $\mathbbm{T}^4
/\mathbbm{Z}_4$ orbifold, in the case that an $S_4$ permutation
symmetry is realized between the states at the four fixed points of
the four dimensional sublattice, the full flavor symmetry will
again be SW$_4$.

For the generic case of the $\mathbbm{T}^4 / \mathbbm{Z}_3$ orbifold we find a
flavor symmetry group $(S_3 \times S_3)\ltimes (\mathbbm{Z}_3)^3$. If the
geometric set-up allows an enlargement of the relabeling symmetry from $S_3
\times S_3$ to $S_9$, this group becomes even larger and can be written as $S_9
\ltimes (\mathbbm{Z}_3)^8$. Similarly, for $\mathbbm{T}^6 /\mathbbm{Z}_3$, the
flavor symmetry could be enhanced to include the permutation symmetry $S_{27}$.

We can also study discrete non-Abelian flavor symmetries at
enhancement points in 6D $\mathbbm{Z}_N \times \mathbbm{Z}_M$
orbifold models. Certain classes of $\mathbbm{Z}_N \times
\mathbbm{Z}_M$ orbifold models at such enhancement points
\cite{Chun:1990iw} are equivalent to Gepner models
\cite{Gepner:1987vz}. Therefore, our analysis is available to such
types of Gepner models.

\section{Comments on symmetry breaking}
\label{sec:symmetrybreaking}

An important question concerns the symmetry breaking patterns of the above
flavor symmetries. As mentioned, the symmetries are explicitly broken by the
introduction of non-trivial discrete Wilson lines. That is, in this case the
degeneracies of mass spectra on different fixed points are lifted, and no
non-Abelian subgroups remain.

On the other hand, when one scalar field in a multiplet of a non-Abelian symmetry
acquires a VEV, a non-Abelian subgroup remains unbroken. Let us consider, for
example, the 2D $\mathbbm{T}^2/\mathbbm{Z}_3$ orbifold. It has
$\Delta(54)=S_3 \ltimes (\mathbbm{Z}_3 \times \mathbbm{Z}_3)$ flavor structure,
and degenerate matter fields at three independent fixed points correspond to a
triplet. Suppose that a scalar field at one of the fixed points, e.g.\
$(\theta,0)$, develops a VEV.\footnote{Giving VEVs to certain scalar fields has
a deep geometrical interpretation in terms of blowing up of the orbifold
singularities, i.e.\ moving in moduli space from the orbifold point to certain
classes of Calabi-Yau manifolds (see \cite{Hamidi:1986vh,Dixon:1986qv} for the
case of standard embedding).} In this case, $S_3$ of $S_3 \ltimes (\mathbbm{Z}_3
\times \mathbbm{Z}_3)$ is broken as $S_3 \rightarrow S_2$, and $(\mathbbm{Z}_3
\times \mathbbm{Z}_3)$ is broken as $\mathbbm{Z}_3 \times \mathbbm{Z}_3
\rightarrow \mathbbm{Z}_3$. Hence, the remaining flavor symmetry is the $D_3 =
S_2 \ltimes \mathbbm{Z}_3$ symmetry, which consists of the 6 elements
\begin{equation}
\left(
\begin{array}{cc}
1 & 0 \\
0 & 1
\end{array}
\right) , \quad
\left(
\begin{array}{cc}
0 & 1 \\
1 & 0
\end{array}
\right) , \quad
\left(
\begin{array}{cc}
\omega & 0 \\
0 & \omega^2
\end{array}
\right) ,\quad
\left(
\begin{array}{cc}
\omega^2 & 0 \\
0 & \omega
\end{array}
\right) ,\quad
\left(
\begin{array}{cc}
0 & \omega  \\
 \omega^2 & 0
\end{array}
\right) ,\quad
\left(
\begin{array}{cc}
0 & \omega^2  \\
 \omega & 0
\end{array}
\right)\; .
\end{equation}
Here, the $\theta$-twisted states $|(\theta,e_1) \rangle$ and
$|(\theta,2e_1)\rangle$ correspond to a $D_3$-doublet. The $D_3$ symmetry is the
only non-Abelian group obtained from $\Delta(54)$ by a VEV of the ${\bf 3}$ as
shown in appendix \ref{app:Delta54}. Recall that only triplets as well as
trivial singlets appear as string fundamental states. However, products of
triplets include other non-trivial representations. If condensates of such modes
form, this could give rise to other breaking patterns. In appendix
\ref{app:Delta54}, all subgroups of $\Delta(54)$ are shown.

Similarly, we can study breaking patterns of the flavor symmetry $S_7 \ltimes
(\mathbbm{Z}_7)^6$, which appears in $\mathbbm{Z}_7$ orbifold models. A similar
type of breaking would lead to the flavor symmetry, $S_n \ltimes
(\mathbbm{Z}_7)^{n-1}$ with $n < 7$.

Let us now comment on how frequent non-Abelian discrete flavor symmetries arise
in realistic orbifold models. The most obvious possibility to accommodate the
observed repetition of families is to construct a model where some or all
families stem from equivalent fixed points. As we have seen, this leads to a
permutation symmetry which, when combined with the other (stringy) symmetries,
gives rise to a non-Abelian flavor symmetry. This symmetry is exact at the
orbifold point, where the expectation values of all (charged) zero modes vanish.
However, the orbifold point is, in general, not a valid vacuum of the model
because of Fayet-Iliopoulos $D$-term. Cancellation of the $D$-term requires
certain fields (which have to be SM singlets for the model to be realistic) to
acquire VEVs. These VEVs lead generically to a spontaneous breakdown of the
non-Abelian discrete flavor symmetries. Another common feature of realistic
orbifold models seems to be the existence of vector-like pairs of SM representations
and anti-representations (cf.\ \cite{Buchmuller:2005jr,Buchmuller:2006ik}). The
SM families mix with these states so that the chiral states observed at low
energies are linear combinations of the states transforming under the
non-Abelian discrete flavor symmetries with flavor singlets (while orthogonal
linear combinations of vector-like matter get large masses). This mixing might
be important in order to reproduce the observed flavor pattern
\cite{Asaka:2003iy,Buchmuller:2006ik}. Altogether, we expect non-Abelian
discrete flavor symmetries to be generic to realistic orbifold models. These
symmetries are usually spontaneously broken in the vacuum, and the pattern of
observed Yukawa couplings is also affected by the mixing of the chiral SM
representations with vector-like states.

We also expect these symmetries to play an important role in understanding the
structure of soft supersymmetry breaking terms.  For example, degeneracy due to
non-Abelian flavor symmetry would be useful to suppress dangerous flavor
changing neutral currents (see, e.g., \cite{Lebedev:2005uh,next}). It appears
possible to arrive at a situation where in the K\"ahler potential, and therefore
in the soft terms, the non-Abelian discrete symmetries survive while the Yukawa
couplings receive important modifications from spontaneous symmetry breakdown.

\section{Conclusions and discussion}

We have studied the origin of non-Abelian discrete flavor symmetries
in string theory.  We have classified all the possible non-Abelian
discrete flavor symmetries which can appear in heterotic orbifold
models. We find that these symmetries exist in many orbifolds, and
have a very simple geometric interpretation. In particular, they are
always present in constructions where the repetition of SM families
is explained by the multiplicity of equivalent fixed points.
A crucial ingredient is the permutation symmetry of
such equivalent fixed points which, together with other symmetries
from the space-group selection rule, generates non-Abelian flavor
symmetries such as $D_4$ and $\Delta(54)$, as well as their direct
products.  A key property of the flavor symmetry is that it is
always larger than the geometrical symmetry of the
compact space. We have  also seen that the flavor
symmetries can get enhanced if the internal space respects certain
symmetries beyond the orbifold twist. We have further discussed how
flavor symmetries can be broken to smaller non-Abelian flavor
symmetries such as $D_3$. At this point, we would like to remind the
reader that the symmetry emerging from the space group is to be
amended by symmetries coming from gauge invariance and $H$-momentum
conservation. That is, there are in general additional gauge factors
(e.g.\ U(1) factors) and discrete $R$-symmetries restricting the
couplings. Besides, from non-Abelian gauge factors one may, in
principle, obtain further non-Abelian discrete symmetries which have
not been discussed here.

It should be interesting to repeat our analysis in non-factorizable orbifolds,
which were recently constructed \cite{Faraggi:2006bs}. Furthermore, our analysis
could be extended to string models with other types of backgrounds, e.g.\ Gepner
manifolds and more general Calabi-Yau manifolds. For example, some of the Gepner
models are equivalent to certain classes of orbifold models at enhancement
points of moduli spaces. On the other hand, blowing-up orbifold singularities
would lead to certain classes of Calabi-Yau manifolds, and such procedure
corresponds to a spontaneous breakdown of (non-Abelian) discrete flavor
symmetries, as discussed before.

In this paper, we have focused on the derivation and classification of
non-Abelian discrete symmetries. It should be interesting to study
phenomenological applications of our results, such as the understanding of the
observed Yukawa matrices of quarks and leptons in terms of spontaneously broken
flavor symmetries, taking into account the mixing with vector-like states.  The
identification of phenomenologically successful flavor symmetries might lead to
the identification of geometries which are particularly useful for obtaining
realistic string compactifications.

\subsection*{Acknowledgments}

It is a pleasure to thank C.~Hagedorn and P.K.S.~Vaudrevange for useful
discussions. T.~K. would like to thank Physikalisches Institut, Universit\"at
Bonn for hospitality during his stay. T.~K.\ is supported in part by the
Grand-in-Aid for Scientific Research \#17540251 and the Grant-in-Aid for the
21st Century COE ``The Center for Diversity and Universality in Physics'' from
the Ministry of Education, Culture, Sports, Science and Technology of Japan.
This work was partially supported by the European Union 6th Framework Program
MRTN-CT-2004-503369 ``Quest for Unification'' and MRTN-CT-2004-005104
``ForcesUniverse''. S.~R.\ received partial support from DOE grant
DOE/ER/01545-869, and would like to thank the Kavli Institute for Theoretical
Physics, Santa Barbara, CA for their hospitality while this paper was being
finished. This research was  supported in part by the National Science
Foundation under Grant No.\ PHY99-07949.

\appendix

\section{Couplings on orbifolds}
\label{app:Couplings}

In this appendix we outline the calculation of coupling strengths on orbifolds.
Let us start with trilinear couplings. Yukawa couplings are obtained by
calculating 3-point functions including three vertex operators corresponding to
massless modes. In heterotic orbifold models, vertex operators consist of a 4D
space-time part, a 6D orbifold part, a gauge part and a bosonized fermion part.
The vertex operators for the 6D orbifold part, the so-called twist fields, are
relevant to our study on flavor symmetries.\footnote{In this appendix, we do not
consider 6D oscillator modes.} One twist field $\sigma_{(\theta^k,\ell)}(z)$ is
assigned to each mode on the fixed point $(\theta^k,\ell)$, that is, each
massless mode corresponding to the boundary condition $(\theta^k,\ell)$. Thus,
Yukawa couplings corresponding to three fields on fixed points
$(\theta^{k^{(j)}},\ell^{(j)})$ $(j=1,2,3)$ are obtained through the calculation of
$\langle \sigma_{(\theta^{k^{(1)}},\ell^{(1)})}
\sigma_{(\theta^{k^{(2)}},\ell^{(2)})} \sigma_{(\theta^{k^{(3)}},\ell^{(3)})}
\rangle$. It can be decomposed as the sum of classical solutions and quantum
fluctuations around them, that is,
\begin{equation}
\left\langle \sigma_{(\theta^{k^{(1)}},\ell^{(1)})}
\sigma_{(\theta^{k^{(2)}},\ell^{(2)})}
\sigma_{(\theta^{k^{(3)}},\ell^{(3)})}  \right\rangle ~=~
Z_\mathrm{qu} \sum_{X_\mathrm{cl}}\, e^{-S_\mathrm{cl}}\;,
\end{equation}
where $Z_\mathrm{qu}$ is the quantum part, $X_\mathrm{cl}$ denotes classical
solutions and  $S_\mathrm{cl}$ is its classical action.
The quantum part $Z_\mathrm{qu}$ is independent of locations of fixed points,
but locations of fixed points are relevant to $S_\mathrm{cl}$.
The classical solution (world-sheet instanton) is obtained as
\begin{equation}
 \partial Z^i_\mathrm{cl}
 ~=~
 a^i\, (z-z_1)^{-1+k^{(1)}v^{(1),i}}\,(z-z_2)^{-1+k^{(2)}v^{(2),i}}\,
 (z-z_3)^{-1+k^{(3)}v^{(3),i}}\;,
\end{equation}
where $Z^i=X^{2i-1}+X^{2i}$ and $z_i$ is the inserted point of $i$-th vertex
operator $\sigma_{(\theta^{k^{(j)}},\ell^{(j)})}$ on the complex coordinate of
the string world-sheet.
The constants $a^i$ are determined by the global monodromy condition, e.g.\ for
the contour around $z_1$ and $z_2$ as
\begin{equation}
 a^i~=~C(k^{(1)}v^{(1),i},k^{(2)}v^{(2),i})(f^{(1),i}-f^{(2),i})\;,
\end{equation}
where $C(k^{(1)}v^{(1),i},k^{(2)}v^{(2),i})$ is a constant depending only on
$k^{(1)}v^{(1),i}$ and $k_{(2)}v^{(2),i}$ and $f^{(a),i}$ denotes the fixed
point corresponding to $\sigma_{(\theta^{k^{(a)}},\ell^{(a)})} $ in the complex
basis $Z^i$. We substitute this solution into the action,
\begin{equation}
 S_\mathrm{cl}~=~\frac{1}{4\pi \alpha'} \int \mathrm{d}^2z\,
 (\partial Z_\mathrm{cl} \bar \partial \bar Z_\mathrm{cl} +
 \bar \partial Z_\mathrm{cl} \partial \bar Z_\mathrm{cl})\;,
\end{equation}
then we can calculate the classical action. Here we take the solution $\bar
\partial Z_\mathrm{cl} =0$. Otherwise, the action does not become finite and does not
contribute to the above amplitude. As a result, the magnitude of Yukawa coupling is
obtained as
\begin{equation}
 Y~\sim~e^{-A},
\end{equation}
where $A$ is the area which the string sweeps to couple. This result is the same
when we use another contour for the global monodromy condition, e.g.\ the
contour around $z_2$ and $z_3$. Note that the fixed point
$(\theta^{k_{(a)}},\ell_{(a)})$ is equivalent to $(\theta^{k_{(a)}},\ell_{(a)} +
(1 -  \theta^{k_{(a)}})\Lambda )$. Thus, we have to sum over classical solutions
belonging to the same conjugacy classes, although the classical solution
corresponding to the shortest distance is dominant and the others lead to larger
classical actions and subdominant effects.

Similarly one can estimate magnitudes of generic $n$-point couplings
\cite{choi}.\footnote{For similar calculations on $n$-point coupling in
intersecting D-brane models see~\cite{Abel:2003yx}.} As before, the
$n$-point function decomposes into a quantum part and a classical part.
Classical solutions have more variety and become complicated. For example,
solutions with $\bar \partial Z  \neq 0$ also lead to %fine finite actions. At
any rate, the classical actions only depend on distances between fixed points
$f^{(a)} - f^{(b)}$ as well as angles between distance vectors, $f^{(a)} -
f^{(b)}$ and $f^{(c)} - f^{(d)}$.

\section{$\boldsymbol{D_4}$ symmetry}
\label{app:D4}

The $D_4$ discrete group has five representations including a doublet $D$, a trivial singlet $A_1$ and three
non-trivial singlets $B_1,B_2,A_2$. Table \ref{tab-4} lists the characters of
these five representations.\footnote{Recall that the character of a group element for
a given representation is defined as the trace of the representation matrix
of the group element (cf.\ \cite{Georgi:1982jb}, chapter 1.13).}

\begin{table}[!htb]
\begin{center}
\begin{tabular}{|c|c|c|c|c|c|}  \hline
Representations & $\mathbbm{1}$ & $-\mathbbm{1}$ & $\pm \sigma_1$ &
$\pm \sigma_3$ & $\mp \I\sigma_2$ \\ \hline \hline
Doublet$-D$ & 2 & --2 & 0 & 0 & 0 \\ \hline
Singlet$-A_1$ & 1 & 1 & 1 & 1 & 1 \\ \hline
Singlet$-B_1$ & 1 & 1 & 1 & --1 & --1 \\ \hline
Singlet$-B_2$ & 1 & 1 & --1 & 1 & --1 \\ \hline
Singlet$-A_2$ & 1 & 1 & --1 & --1 & 1 \\ \hline
\end{tabular}
\end{center}
\caption{Character table for $D_4$.}
\label{tab-4}
\end{table}

A product of two doublets decomposes into four singlets,
\begin{equation}
(D \times D)~=~A_1 + B_1 + B_2 + A_2\; .
\end{equation}
More explicitly, we consider two
$D_4$ doublets $S_A$ and $\bar S_A$ $(A=1,2)$.
Their product $S_A \bar S_B$ can be decomposed
in terms of  $A_1,B_1,B_2,A_2$,
\begin{align}
S_1 \bar S_1 + S_2 \bar S_2 &\sim~ A_1\; ,
&
S_1 \bar S_2 + S_2 \bar S_1 &\sim~ B_1\; ,\nonumber\\
S_1 \bar S_1 - S_2 \bar S_2 &\sim~ B_2\; ,
&
S_1 \bar S_2 - S_2 \bar S_1 &\sim~ A_2\; .
\end{align}

%\newpage

\section{$\boldsymbol{\Delta(54)}$ symmetry}
\label{app:Delta54}

%Here, we list some group-theoretical aspects of $\Delta(54)$.  (See
%also Ref.~\cite{Muto:1998na}.)
Group-theoretical aspects of $\Delta(54)$ can be found
in~\cite{Fairbairn:1964,Muto:1998na}. It is a discrete subgroup of
$\mathrm{SU}(3)$, i.e.\ the group $\Delta(6n^2)$ (with $n=3$) and order $54\, (=3!
\cdot 3^2)$. The generators of $\Delta(54)$ are given by the set
\begin{equation}
 \left(
 \begin{array}{ccc}
 0 & 1 & 0 \\
 0 & 0 & 1 \\
 1 & 0 & 0
 \end{array}
 \right)\;,  \quad
 \left(
 \begin{array}{ccc}
 e^{\I\, \alpha} & 0 & 0 \\
 0 & e^{\I\, \beta} & 0 \\
 0 & 0 & e^{-\I\,(\alpha + \beta)}
 \end{array}
 \right)\;, \quad
 \left(
 \begin{array}{ccc}
 e^{\I\, \alpha} & 0 & 0 \\
 0 &  0 & e^{\I\, \beta} \\
 0 & e^{-\I\,(\alpha + \beta)} & 0
 \end{array}
 \right)\;,
 \end{equation}
with $\alpha = 2 \pi j/3$, $\beta = 2 \pi k/3$ and $j,k$ integers. In general,
it has four three dimensional irreducible representations $\mathbf{3},\,
\mathbf{\bar{3}}, \, \mathbf{3'},\, \mathbf{\bar{3'}}\,$, four two dimensional
ones $\mathbf{2}_{1},\, \mathbf{2}_{2}, \, \mathbf{2}_{3}, \, \mathbf{2}_{4}$
and two one dimensional ones $\mathbf{1}_{1}, \, \mathbf{1}_{2}$. Their
characters are summarized in table \ref{tab:characters1}. The subgroups of
$\Delta(54)$ are given in table \ref{tab:repbreakings}.

\begin{table}[!h]
\centerline{
\begin{tabular}{|c|rrrrrrrrrr|}
\hline
irrep & 1a  & 6a  & 6b  & 3a  & 3b  & 3c  & 2a  & 3d  & 3e  & 3f \\
      & \scriptsize{(1)} & \scriptsize{(9)} & \scriptsize{(9)} & \scriptsize{(6)} & \scriptsize{(6)} & \scriptsize{(6)} & \scriptsize{(9)} & \scriptsize{(6)} & \scriptsize{(1)} & \scriptsize{(1)} \\
\hline
$\mathbf{1}_{1}$ & 1 & 1 & 1 & 1 & 1 & 1 & 1 & 1 & 1 & 1 \\
$\mathbf{1}_{2}$ & 1 & -1 & -1 & 1 & 1 & 1 & -1 & 1 & 1 & 1 \\
$\mathbf{2}_{1}$ & 2 & 0 & 0 & 2 & -1 & -1 & 0 & -1 & 2 & 2 \\
$\mathbf{2}_{2}$ & 2 & 0 & 0 & -1 & -1 & -1 & 0 & 2 & 2 & 2 \\
$\mathbf{2}_{3}$ & 2 & 0 & 0 & -1 & -1 & 2 & 0 & -1 & 2 & 2 \\
$\mathbf{2}_{4}$ & 2 & 0 & 0 & -1 & 2 & -1 & 0 & -1 & 2 & 2 \\
$\mathbf{3'}$ & 3 & $-\bar{\omega}$ & $-\omega$ & 0 & 0 & 0 & -1 & 0 & $3\bar{\omega}$ & $3\omega$ \\
$\mathbf{\bar{3'}}$ & 3 & $-\omega$ & $-\bar{\omega}$ & 0 & 0 & 0 & -1 & 0 & $3\omega$ & $3\bar{\omega}$ \\
$\mathbf{\bar{3}}$ & 3 & $\omega$ & $\bar{\omega}$ & 0 & 0 & 0 & 1 & 0 & $3\omega$ & $3\bar{\omega}$ \\
$\mathbf{3}$ & 3 & $\bar{\omega}$ & $\omega$ & 0 & 0 & 0 & 1 & 0 & $3\bar{\omega}$ & $3\omega$ \\
\hline
\end{tabular}
}
\caption{Character table of the group $\Delta(54)$ (with  $\omega = e^{2 \pi
\I/3}$). The second row gives the number of elements in the certain class and
the second column the dimension of the representation.}
\label{tab:characters1}
\end{table}

%\vspace{2cm}
%\enlargethispage{1cm}

\begin{table}[!h]
\centerline{
\begin{tabular}{|c|r|c|c|r|}
\cline{1-2}\cline{4-5}
subgroup & decomposition of $\boldsymbol{3}$  & & 
subgroup & decomposition of $\boldsymbol{3}$\\
\cline{1-2}\cline{4-5}
\cline{1-2}\cline{4-5}
$\Delta(27)$ & $\boldsymbol{3}$ $\phantom{I^{\phantom{\int^{I}}}_{\phantom{\int_{I}}}}$ 
& &
$S_3$ & $\boldsymbol{2} + \boldsymbol{1_1}$ $\phantom{I^{\phantom{\int^{I}}}_{\phantom{\int_{I}}}}$ 
\\
$S_3\ltimes \mathbbm{Z}_{3}$ & $\boldsymbol{2_2} + \boldsymbol{\overline{1_4}}$ $\phantom{I^{\phantom{\int^{I}}}_{\phantom{\int_{I}}}}$ 
& &
$\mathbbm{Z}_2$ & $2 \cdot \boldsymbol{1_1} + \boldsymbol{1_2} $  $\phantom{I^{\phantom{\int^{I}}}_{\phantom{\int_{I}}}}$
\\
$\mathbbm{Z}_{3} \times \mathbbm{Z}_{3}$ & $\boldsymbol{\overline{1_2}} + \boldsymbol{\overline{1_3}} + \boldsymbol{1_4}$ $\phantom{I^{\phantom{\int^{I}}}_{\phantom{\int_{I}}}}$ 
& &  &\\
\cline{1-2}\cline{4-5}
\end{tabular}
}
\caption{Decomposition of the $\boldsymbol{3}$ of $\Delta(54)$ into irreps of the particular subgroup under the breaking.}
\label{tab:repbreakings}
\end{table}

\end{document}